\documentclass{pasa}%

\title[One year of monitoring the Vela pulsar using a Phased Array Feed]{One year of monitoring the Vela pulsar using a Phased Array Feed}
\author[Sarkissian et al.]{John M. Sarkissian$^1$, John E. Reynolds$^2$, George Hobbs$^2$, Lisa Harvey-Smith$^2$ \\
\affil{$^1$CSIRO Astronomy and Space Science, Parkes Observatory, PO Box 276, Parkes NSW 2870, Australia}%
\affil{$^2$CSIRO Astronomy and Space Science, Australia Telescope National Facility, PO Box 76, Epping NSW 1710, Australia}}%
\jid{PASA}
\doi{10.1017/pas.\the\year.xxx}
\jyear{\the\year}

\usepackage[authoryear]{natbib}
\bibpunct{(}{)}{;}{a}{}{,}
\usepackage{aas_macros}
\usepackage{hyperref} 
\hypersetup{colorlinks,citecolor=blue,linkcolor=blue,urlcolor=blue}

\begin{document}%
\begin{abstract}
We have observed the Vela pulsar for one year using a Phased Array Feed (PAF) receiver on the 12-metre antenna of the Parkes Test-Bed Facility.  These observations have allowed us to investigate the stability of the PAF beam-weights over time, to demonstrate that pulsars can be timed over long periods using PAF technology and to detect and study the most recent glitch event that occurred on 12 December 2016.  The beam-weights are shown to be stable to 1\% on time scales on the order of three weeks.  We discuss the implications of this for monitoring pulsars using PAFs on single dish telescopes.
\end{abstract}
\maketitle%
\section{Introduction}
\label{sec:intro}

The Parkes Test-Bed Facility (PTF) was commissioned in 2009 as a test platform for the Commonwealth Scientific and Industrial Research
Organisation (CSIRO) program of developing Phased Array Feed (PAF) receiver technology. We have conducted regular monitoring of the Vela pulsar and report here on our observations and on the stability of the PAF system. Our data set provides the first long-term, timing observations of a pulsar using a PAF.  We make all our observational data publically available.

PAFs are now being used for a variety of observations on different telescopes.   Some of the first observations of a PAF were carried out with the Westerbork telescope (van Cappellen, Bakker \& Oosterloo 2009a\nocite{2009isap.confE...1V} \& 2009b\nocite{5171752}).  They reported on the simultaneous detection of the two bright, northern pulsars, PSRs~B0329+43 and B0355+54.  The first mainstream astronomy papers using a PAF came from the Boolardy Engineering Test Array (BETA) which consisted of six 12-m antennas situated in the Murchison Radio-astronomy Observatory (MRO) incorporating the first generation, Mark-I PAF (Hotan et al. 2014\nocite{2014PASA...31...41H}).  This system was used to observe the intermittent pulsar~PSR~J1107$-$5907 (Hobbs et al. 2016\nocite{2016MNRAS.456.3948H}), to produce wide-area continuum images (e.g., Serra et al. 2015\nocite{2015MNRAS.452.2680S}, Heywood et al. 2016\nocite{2016MNRAS.457.4160H}) and for spectral line observations (e.g., Allison et al. 2015\nocite{2015MNRAS.453.1249A}, Harvey-Smith et al. 2016\nocite{2016MNRAS.460.2180H}).  The Australian Square Kilometre Array Pathfinder (ASKAP) telescope is currently being commissioned.  When complete, it will consist of 36 12-m antennas each incorporating a second-generation (Mark-II) PAF. The primary science goals for this telescope are described by Johnston et al. (2007\nocite{2007PASA...24..174J}) and Johnston et al. (2008\nocite{2008ExA....22..151J}).   One of these Mark-II PAFs was recently installed on the Parkes 64-m antenna and was used to study pulsar and transient observations.  For all these science and engineering projects, the PAF has been used for periods of a few days and the stability of PAFs over long time-scales has not been studied in detail. The driving motivation therefore, for our work has been to investigate the stability of these PAFs on time scales of several weeks or more, and to demonstrate their utility for long-term observing programmes.

A BETA Mark-I PAF was installed on a testbed system at the Parkes observatory site with the primary goal of refining the design of the PAF for the full ASKAP system. We observed the Vela pulsar (PSR~J0835$-$4510) on a close-to daily cadence to study the stability of the PAF over long durations and to study the long-term timing irregularities that Vela is known to exhibit.

Vela is a well-studied pulsar.  It is the brightest pulsar in the radio observing band and many telescopes carry out regular monitoring observations - including the 26m telescope at Hartebeesthoek in South Africa (e.g., Buchner 2013\nocite{2013IAUS..291..207B}), the 14m and 26m telescopes at Mt Pleasant in Tasmania (e.g., Dodson et al. 2007\nocite{2007Ap&SS.308..585D}, Palfreyman et al. 2016a\nocite{2016ApJ...820...64P} respectively) and, more recently, the Molonglo telescope.  It is also commonly observed with the 64-m Parkes radio telescope. Recently Shannon et al. (2016\nocite{2016MNRAS.459.3104S}) characterised 21\,years of Parkes observations.  They provided updated parameters for eight glitch events (a glitch event is where the pulse frequency suddenly changes) and the observed timing noise (a phenomena that leads to low-frequency noise in the timing residuals).     In \S2 of this paper we describe our observations of the Vela pulsar. In \S3.1 we discuss the stability of the beam-weights on long time scales and in \S3.2 we present our long-term monitoring of the Vela pulsar.  We conclude in \S4.

\section{Observations}

\begin{table}\begin{center}
\caption{Observations of the Vela pulsar using the PAF on the PTF.}
\begin{tabular}{p{3.9cm}p{4cm}}
\hline
Parameter & Value\\
\hline
Telescope diameter (m)& 12\\
Number of PAF beams & 1 \\
Approximate system temperature (K) & 50\\
Frequency range (MHz) & 1191.5 -- 1207.5\\
\hline
Observation span & 2016/02/09 -- 2017/02/26 \\
Number of observations & 502 \\
Observation length (min) & 20 \\
Largest data gap (d) & 14 \\
Median error bar size ($\mu$s) & 6.4 \\
\hline
\end{tabular}
\end{center}\end{table}

The PTF comprises a 12-metre fully-steerable antenna made by Patriot
Antenna Systems, with the first of the ASKAP Mark-I PAF receivers mounted
at prime focus (focal ratio = 0.4). The PAF is of the chequerboard design (Hay and
O'Sullivan 2008\nocite{RDS:RDS5546}) and has 188 individual ports which sample the electric field
in the focal plane in two orthogonal linear polarizations. The analogue
signals are transported to an adjacent building where they are digitized and
combined in digital beam-formers to form up to nine dual-polarization
individually-steerable beams. The PAF, digitizers and beam-formers are
identical to those used for the BETA (Hotan et al. 2014). 

Although the system has not been designed explicitly for high-time resolution
astronomy, we make use of built-in diagnostic tools that allow limited
capture of beam-formed voltage (base-band) time series. Sixteen
contiguous frequency channels, each of 1MHz bandwidth, can be captured
continuously in ~1-second bursts, with a dead-time of several seconds for
read-out, the length of which depends on the number of captured beams.
For most of these observations we captured a single dual-polarized beam,
sampled for 0.884736 seconds every 6 seconds. Each individual 1MHz datastream is over-sampled at the rate of 1.185MHz
(1MHz x 32/27). This over-sampling was required for the 2-stage
polyphase filterbank in the Mark-I (BETA) design for ASKAP, although the
second stage filterbank is not implemented on the PTF. The relatively low duty cycle
restricts the maximum pulsar period that can usefully be observed but
poses no problem for young or millisecond pulsars.

The voltage data is written as VDIF (VLBI Data Interchange Format) files
and subsequently converted into time-averaged and folded data in PSRFITS
format (Hotan, van Straten \& Manchester 2004\nocite{2004PASA...21..302H}), allowing standard pulsar processing tools to be
used.  In the top part of Table~1 we summarise the properties of the observing system.  

Observations of the Vela pulsar (PSR~J0835$-$4510) were performed daily,
depending on the availability of the PTF. Each observation 
typically lasted 20 minutes of which 0.884736 seconds of every six
seconds would contain actual data.  This is equivalent to 176.9472
seconds of data overall for each observation. Immediately following an observation, the
output VDIF files are converted to PSRFITS format using software
specifically written for this project.  In the bottom part of Table~1 we summarise our observations of the Vela pulsar.

\section{Results and Discussion}

\subsection{Beam-weight stability}

\begin{figure}

\includegraphics[angle=-90,width=8cm]{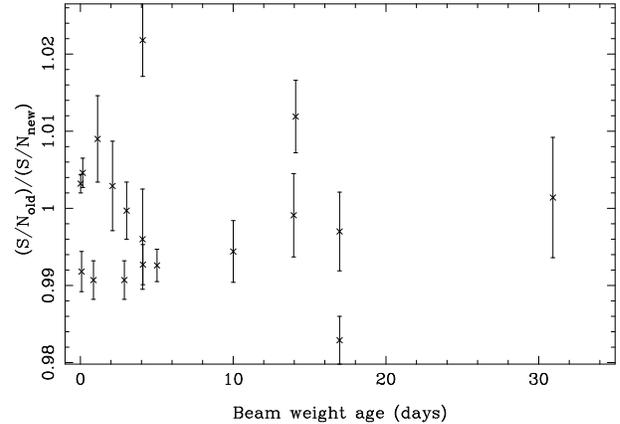}
\caption{On-axis (boresight) response of beam-weights as a function of their age plotted as the ratio of the S/N of the pulse profile using the original weights to the S/N obtained with recent weights.  Error bars are standard errors in the mean after averaging the 16 independent 1\,MHz frequency channels. It is clear that the beam-weights are stable to 1\% over a period of order 3 weeks. }\label{fg:beamWeights}
\end{figure}

It has already been shown that PAFs can be used to observe astronomical sources.  PAF beams are formed by weighting the signal from the individual elements on the PAF.  The required weighting is known to change with time (through ageing or instability of the electronic components and because of temperature variations) and yet, until now, there has not been a comprehensive study of how long a given set of beam-weights can be used without significant degradation of the astronomical signal.   

We have investigated beam-weight stability by making a number of observations of the Vela pulsar with two simultaneous beams, both centred on the pulsar, with one beam formed from beam-weights determined shortly before the observation, and the second from beam-weights determined several days previously. We then compared the signal-to-noise ratio (S/N) of the folded pulsar profile in the two beams. Figure~\ref{fg:beamWeights} shows for each of these observations the measured S/N of the older beam with respect to (divided by) that of the newer beam as a function of the interval between the two beam-weight solutions. As the data recorded in each beam should be identical except for the different beam-weights any ageing effects should appear as a decline in this quotient over time. No such trend is visible in our data, out to intervals of two to four weeks. It is clear from figure 1, that the beam-weights are stable to 1\% over a period of order three weeks. The paucity of data at intervals longer than $\sim$15 days is  caused by unscheduled outages in the mains power supply which typically required a complete restart of the system every 10 to 20 days, introducing random changes in delay between PAF elements.

The process of computing beam-weights and of taking the observations, treats each frequency channel separately so each channel provides an independent measurement of the beam-weight ageing. In the figure we have therefore plotted the mean over all 16 channels with the error bars indicating the standard error in the mean. However, the indicated errors do not reflect systematic effects common to all channels and therefore underestimate the true uncertainty in each  measurement, as can be seen from the scatter between plotted points. There is also a suggestion of structure in the plotted results, with an apparent cluster of points around ordinate 0.99 for T $<$ 6 days, but we can identify no obvious physical explanation for this and suggest this is simply statistical noise.

The technique here described is sensitive only to changes of gain or phase from one PAF element to another, and not to changes that are uniform across the array. It is also not sensitive to small relative changes of gain and phase between elements as these do not affect the sensitivity (SNR) on the bore-sight to first order. However, the results do give us confidence that the useful life of a beam-weight solution for this PAF is measured in days or weeks rather than hours or minutes, without any recalibration.

\subsection{Timing of the Vela pulsar}

\begin{figure}
\includegraphics[width=8cm]{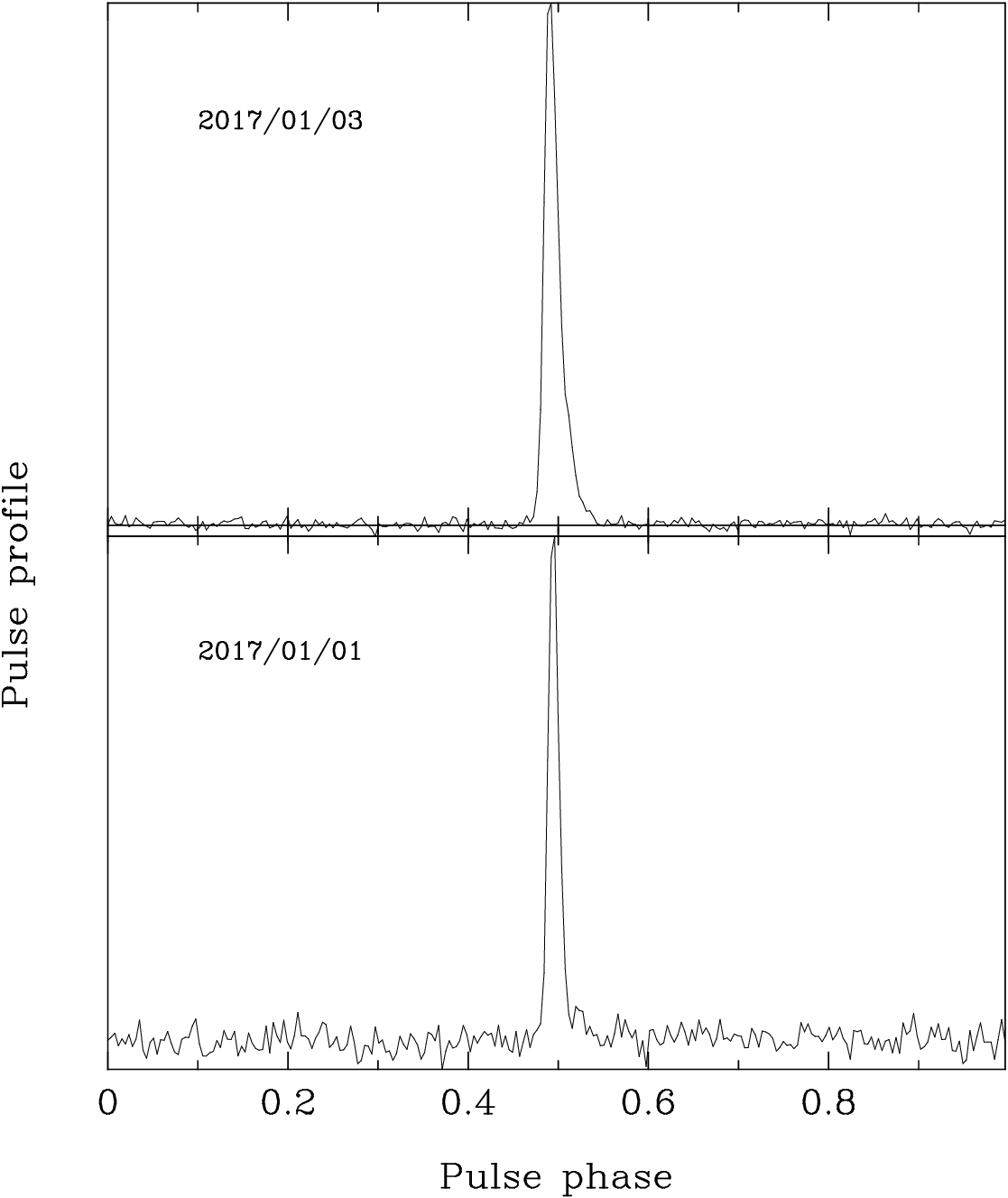}
\caption{Typical pulse profiles of the Vela pulsar achieved with the PTF. The top plot shows a high S/N profile from 3 January 2017 and the bottom plot a more typical pulse profile obtained on 1 January 2017.}
\end{figure}

In order to demonstrate the data quality that we can achieve with the PTF we show, in Figure~2, a typical pulse profile of the Vela pulsar obtained on 1 January 2017 and a high S/N profile 3 January 2017.  Clearly the signal-to-noise of the profiles are adequate for determining pulse arrival times.  We formed pulsar timing residuals using the \textsc{tempo2} (Hobbs, Edwards \& Manchester 2006\nocite{2006MNRAS.369..655H}) software package with an initial timing model developed for young-pulsar monitoring experiments with the Parkes 64-m telescope (Kerr et al. 2014\nocite{2014atnf.prop.6317K}).    We have compared our residuals with those from the University of Sydney's Molonglo telescope (Jankowski, F., private communication) and the University of Tasmania's 26m telescope at Mt Pleasant (Palfreyman, J., private communication) and find that they are in agreement. The timing in the PTF system is accurate at the microsecond level (the same system was used to record VLBI data and no timing problems have been identified).   The pulse TOAs were determined using a standard profile template that was derived from the observation with the highest S/N ratio.  The template is available in the online data collection and can be used to determine the absolute phase of our measurements with respect to other data sets.

\begin{figure}
\includegraphics[angle=-90,width=8cm]{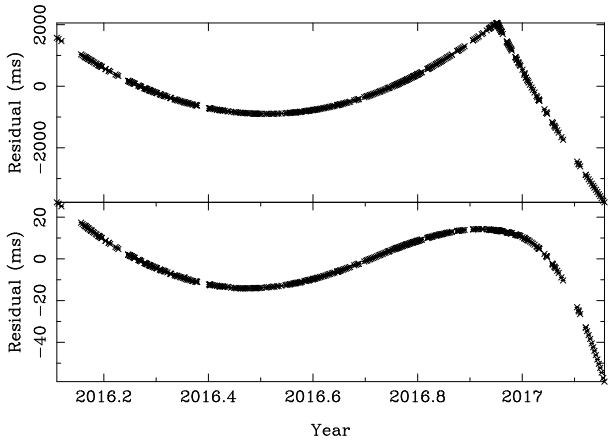}
\caption{The full data set showing the glitch event (top); the observed timing residuals after removing a fixed frequency and time derivative only (upper panel), and after fitting for the glitch event (lower panel).}\label{fg:residuals}
\end{figure}

The resulting timing residuals (after only fitting for the pulsar's pulse frequency and its first time derivative) are shown in the top panel of Figure~\ref{fg:residuals}.  Note that these residuals have a range much greater than one pulse period.  In order to obtain a phase connected solution we have used the pulse numbering scheme within the \textsc{tempo2} software package that enables the user first to whiten the data, then to allocate a pulse number to each individual arrival time and then finally to plot the residuals without any whitening scheme.  The uncertainties on each arrival time are smaller than the symbol used in the Figure. It is therefore clear that small telescopes such as the PTF are sufficient for long-term monitoring of pulsars such as Vela.  It is also clear that a glitch event occurred near the end of 2016.

This event was quickly reported by Palfreyman (2016b\nocite{2016ATel.9847....1P}) at MJD 57734.48  (12th December 2016) between 11:31
and 11:46 UTC, with $\Delta \nu/\nu = 1.3\times10^{-6}$. Our first observation was made independently, less than one hour after the glitch at 12:34 UTC and the previous observation was at 11 December 2016 at 21:10 UTC.   We used the glitch plugin in \textsc{tempo2} to both visualise and to obtain an initial parameterisation for the glitch event.  We were able to fit for the epoch of the glitch and the changes in the pulse frequency and its derivative using $\sim$20\,d of observations either side of the glitch.  The glitch plugin also clearly indicated a decaying term, which we modelled as an exponential recovery.  This is a non-linear fit and we were unable to get a standard \textsc{tempo2} fit to converge.  In order to measure the time constant we therefore trialled a range of values and selected the time constant that led to the smallest rms residual.  The resulting glitch parameters are listed in Table~\ref{tb:glitchParams}.

\begin{table}\begin{center}
\caption{Parameters for the 12 December 2016 Vela glitch.  Uncertainties are given in parenthesis and refer to the last significant digit. The top five parameters are the determined values. The bottom three parameters are derived values.}\label{tb:glitchParams}
\begin{tabular}{ll}
\hline
Parameter & Value \\
\hline
$\phi_g$ & $-$0.0077(5) \\
$\Delta \nu_g$ (Hz) & $1.6044(2) \times 10^{-5}$ \\
$\Delta \dot{\nu_g}$ (s$^{-2}$) & $-1.21(3) \times 10^{-13}$ \\
$\nu_d$ (Hz) & $1.29(8) \times 10^{-7}$ \\
$\tau_d$ (d) & 0.96(17) \\
\\
Glitch epoch (MJD) & 57734.4855(4) \\
$\Delta \nu/\nu$ &  $1.43387(2) \times 10^{-6}$ \\
$\Delta \dot{\nu}/\dot{\nu}$ & 0.0077(2) \\
\hline
\end{tabular}
\end{center}
\end{table}

The timing residuals after modelling the glitch using the parameters listed in the Table are shown in the bottom panel of Figure~\ref{fg:residuals}.  Our glitch epoch is consistent with those initially presented in Palyfreyman (2016b) and here we present a more complete parameterisation of the glitch event.  The ATNF pulsar glitch catalogue\footnote{\url{http://www.atnf.csiro.au/people/pulsar/psrcat/glitchTbl.html}} lists 19 glitches for the Vela pulsar, beginning with the very first recorded glitch of a pulsar in 1969 (Radhakrishnan \& Manchester 1969\nocite{1969Natur.222..228R}). Typical $\Delta \nu/\nu$ values are $\sim 10^{-6}$ and $\dot{\nu}/\nu \sim 0.01$.  The glitch event reported here is therefore typical in size of previous Vela glitches.  Many of the Vela glitches have recovery events.  The time scales listed in the catalogue range from fractions of a day to almost a year (McCulloch et al. 1990\nocite{1990Natur.346..822M}). This is likely to be related to how close the subsequent observations were to the actual glitch event.  Our multiple observations within days of the glitch has enabled us to identify the $\sim 1$\,d time constant reported here. In order to measure the time constant we therefore trialled a range of values and selected the time constant that led to the smallest rms residual.  The output of this grid search is shown in Figure 4.  We then chose the time constant leading to the smallest rms residual (of 0.98\,d) and included that in the \textsc{tempo2} parameter fit.  This time the fit converged and gave a time constant of 0.96(17)\,d.  The resulting glitch parameters are listed in Table 2.

\begin{figure}
\includegraphics[angle=-90,width=8cm]{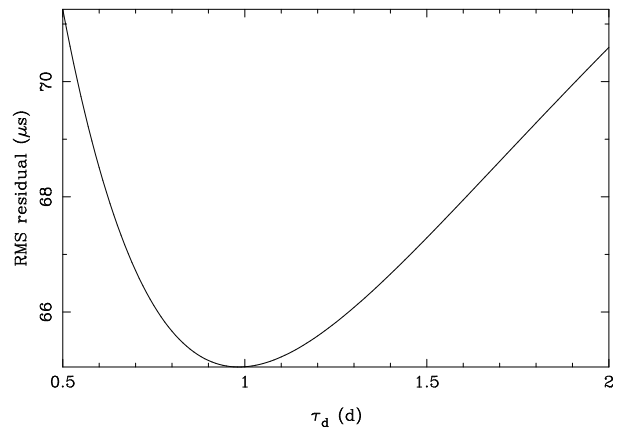}
\caption{The rms timing residual obtained using a grid search for different glitch time decay constants.  The smallest value occurs at 0.98 days.}
\end{figure}

\subsection{Data access}

The folded data taken for this project (along with our processing scripts) are available without any embargo period from the CSIRO data access portal (DAP; \url{data.csiro.au}; Hobbs et al. 2011).  The data collection (available from \url{http://doi.org/10.4225/08/58eb71cd397f3}) is divided into three directories as follows:
\begin{itemize}
\item paper\_files: This directory contains all the fold-mode observation files in the PSRFITS format.  These files can be viewed and processed using the \textsc{PSRCHIVE} software package.
\item paper\_parTim: This directory contains the parameter files (with file extensions .par) and arrival time files (with file extensions .tim) along with auxiliary files that were used to select regions around the glitch event etc.   The parameter and arrival time files can be processed as normal using \textsc{tempo2}.  We also include a simple script (getTd) that was used to carry out a grid-search over the glitch decay timescale.
\item paper\_scripts: Here we provide \textsc{getToAs} that carries out the processing steps from the raw data to pulse arrival times and \textsc{processData} that includes the processing steps for processing creating a timing solution and modelling the glitch.
\end{itemize}

We encourage the use of these files for other analyses of the Vela pulsar, but remind the user that these observations were obtained using a test-bed facility in a non-standard observing mode. 

\section{Conclusion}
We have investigated and confirmed the stability of PAF beam-weights over time scales of several weeks and we have demonstrated that pulsars can be timed on year-long time scales using this PAF technology.  The wide-field-of-view and relatively wide bandwidths achievable using PAFs suggest that such receivers will become more common in the near future for main-stream radio astronomical observations.

The Mark-II PAF that was recently installed on the 64-m antenna at Parkes, has been shipped to Germany for installation on the Effelsberg 100-m telescope. In addition, Jodrell Bank Observatory, Green Bank, FAST and many more observatories
are also considering the installation of PAFs at their telescopes. Our investigations have been extremely useful for these, since they provided the opportunity to better understand and characterise the evolving PAF technologies.

The work we have performed with the PTF PAF has shown that it is a satisfactory instrument for monitoring bright pulsars. We will continue to monitor the Vela pulsar and will also use the PTF to carry out an in-depth study of bright emission modes for intermittent pulsars such as PSR  J1107-5907 (Young et al. 2014\nocite{2014MNRAS.442.2519Y}).

In the final years of the close monitoring of the Vela pulsar by Dodson et al. (2007), the data rate was increased such that the pulse times of arrival (TOAs) could be measured every 10 seconds. For the observations here described, it was not possible to do this owing to the limitations imposed by the current system. However, it is a possibility that is under serious consideration for the future, with a new receiver and/or backend on the 12m PTF.

\begin{acknowledgements}
The Parkes Test-Bed Facility is part of the Australia Telescope, which is funded by the Commonwealth of Australia for operation as a National Facility managed by
the Commonwealth Scientific and Industrial Research
Organisation (CSIRO). We thank the staff of the CSIRO Parkes Observatory for their invaluable
efforts in commissioning and maintaining the PTF and to Aidan Hotan
for his initial investigations on using the PTF for observing pulsars.  We thank Jim Palfreyman and Fabian Jankowski for allowing us to compare our Vela observations with theirs.  We thank Dick Manchester and Jimi Green for providing useful comments on our work.
\end{acknowledgements}

\bibliographystyle{pasa-mnras}
\bibliography{js_references_2}
\end{document}